\documentclass[12pt,a4paper]{article}
\usepackage{amssymb,amsmath,amsthm}
\usepackage[english]{babel}
\usepackage{amsfonts,amssymb}
\usepackage{latexsym}
\begin{document}
\large

\begin{center}
{\bf Investigation of the kinetic equation of cascade
fragmentation theory at not self-similar subdivision}

Brodskii R.E., Virchenko Yu.P.
\end{center}

{\bf Introduction.} In this work we
investigate the cascade fragmentation process of material sample.
This process is the particular case of the cascade kinetic
processes which arise in different fields of theoretical physics (see, for
example, \cite{Bar}).
In spite of the fact that cascade type of fragmentation is the simplest one
and the fragmentation which has no  the cascade character is more widespread
in nature, such processes are important from the physical point of view
and have been not completely studied.
In particular, it is not studied the cascade fragmentation process that
is not completely self-similar in contrast to the analyzed one in
the classical work \cite{Kolm}.
In the present communication, we investigate  the
fragmentation being not self-similar
that it is analogous to the process introduced in the paper \cite{Our}.

{\bf General equation.} We describe the material fragmentation process by
the particle density $n(r, t)$.
The value $n(r, t) dr$ is the average fragment number which
have their sizes in the interval $(r, r+dr)$ at the time moment $t$.
The fragmentation process is determined completely by the conditional
probability density $P(\rho, r, t)$ of the splinter formation of
given fragment. More strictly, the probability of the formation of
splinter that has the size in the interval
$(\rho, \rho + d \rho)$ during the time interval $(t, t+dt)$ is equal to
$P(\rho, r, t)d\rho dt$ when it is occurred the decay of the fragment with
the size $r$. We introduce also the intensity $\mu(r, t)$ of the fragment
decay. It is the average number of decays of fragments with the size $r$
in the time interval $(t, t+dt)$.

The evolution of the density $n(r,t)$ is described in terms of
values pointed out and on the base of the kinetic equation
$$
\dot n(r,t) = \int\limits_r^\infty P(r, \rho, t) n(\rho, t) d\rho -
\mu(r,t)n(r,t)\,. \eqno(1)
$$
In the article \cite{OurKolm},
it has been studied the fragmentation process
with the scale-invariant subdivision having temporally independent
conditional probability density of transitions
$$
P(\rho, r, t) = \frac 1 r P(\rho/r)\,.
$$
It has been done following the idea contained in
the pioneer work \cite{Kolm}.
At the present communication, it is considered the fragmentation process
that has no the scale-invariant subdivision when the transition
probability density has the form
$$
P(\rho, r, t) = P(\rho/r) \eqno(2)$$
as it has been done in the paper \cite{Our}. So, the chosen subdivision
scale exists since, in this case, the density $P(\rho, r, t)$ has the
physical dimensionality of the inverse length.
Following to works \cite{OurKolm}, \cite{Yan}, we take into
account the volume conservation in the fragment system
$$
V = \int\limits_0^\infty r^3 n(r, t)dr = {\rm const}\,. \eqno(3)
$$
This property gives the next relation for the intensity
$$
\mu(r,t)=\int\limits_0^r P(\rho/r)\left(\frac{\rho}{r}\right)^3  d \rho =
r \int\limits_0^1 P(x) x^3 dx \equiv \mu r, \quad \mu = {\rm const}
\,.\eqno (4)
$$
Thus, we consider the next kinetic equation for the particle density
$$
\dot n(r,t) = \int\limits_r^\infty P(r/\rho) n(\rho, t) d\rho -
\mu r\, n(r,t)\,. \eqno(5)
$$

{\bf The limit distribution.}
The formula (2) permits to use the Mellin transformation
$$
M(s,t) = \int\limits_0^\infty r^{s-1}n(r, t)dr
$$
for study some solutions of the equation (5).
In this case,
$M(1,t) = N(t)$ is the average fragment number at the moment $t$;
$M(2,t) = N(t)\lambda(t)$ where $\lambda(t)$ is the average fragment size;
$M(3,t) = S(t)$ is the total fragment surface area;
$M(4,t) = V = {\rm const}$ is the total fragment volume.

The application of the Mellin transformation
to both parts of the equation (5) leads to the equation
$$
\dot M(s,t) = M(s+1, t)(p(s) - \mu), \eqno (6)
$$
where $p(s) = \int\limits_0^1 x^{s-1}P(x)dx$ and $\mu = p(4)$ according to
(4).
In particular,
$$
M(3,t) = M(3, 0) + V(p(3)-p(4))t \eqno (7)
$$
since $M(4, t) = V = {\rm const}$. The asymptotic behavior
of the distribution $f(r, t)$ density at $t \to \infty$ is of interest for
us. In this case, it is sufficient to use the second term
$M(3,t) \sim V(p(3)-p(4))t$ only in the last formula.
From equations (6) and (7), we find
$$
M(2,t) \sim  \frac 1 2 V(p(3)-p(4)) (p(2)-p(4)) t^2 = N(t)\lambda(t),
\eqno (8)
$$
$$
M(1, t) \sim \frac 1 6 V(p(3)-p(4)) (p(2)-p(4)) (p(1)-p(4)) t^3  = N(t)\,.
\eqno (9)
$$
Therefore, it follows
$$
\lambda(t) \sim \frac{M(2,t)}{M(1,t)} = \frac{3}{p(1)-p(4)}\frac 1 t\,.
\eqno(10)
$$

{\bf The analysis of the model $P(x) \sim x^\alpha$.}
In general case, the study of solutions of the differential equation (6)
is sufficiently difficult. However, for the probability
distribution density $f(r,t) = n(r,t)/N(t)$
on fragment sizes when the variable $r$ is replaced by
$\lambda(t) r$, the analysis of this equation becomes much more simple.
Let us introduce the designation by $F(s,t)$ of the Mellin transformation
$$
F(s, t) = \lambda(t) \int\limits_0^\infty r^{s-1}f(\lambda (t)r,t)dr\,.
$$
of the distribution $\lambda(t)f(\lambda (t)r, t)$ density.
Changing the variables $r\Rightarrow \lambda(t)r$ in
the equation (5) and computing the Mellin transformation pointed out,
we obtain
$$
\dot F(s,t) = - F(s,t) \frac d {dt} \ln \left(\frac N \lambda
\right) - \frac s \lambda
F(s,t)\dot \lambda + \lambda (p(s) -\mu)F(s+1, t)\,. \eqno (11)
$$
The distribution $\lambda (t)f(\lambda (t)r, t)$ density and
its Mellin transformation $F(s,t)$ are tended to $f_\infty (r)$ and
$F(s)$ at $t \to \infty$, correspondingly. These limit functions
are nontrivial ($\ne 0; \infty$). In this case,
$\dot F(s,t) \to 0, \quad t \to \infty$. So, the equation (11) take the form
$$
0 =  - F(s) \frac d {dt} \ln \left(\frac N \lambda
\right)+\lambda (p(s)-\mu)F(s+1) -
\dot\lambda \frac s \lambda  F(s)\,.  \eqno(12)
$$
at asymptotically large values $t$.
Let us find the limit distribution density $f_\infty (r)$ for the
model dependence $P(x)$,
$$
P(\rho/r) = C \left(\frac \rho r \right)^\alpha\,,\quad
\alpha > 0 \,. \eqno(13)
$$
In this case,
$$
p(s) = \frac C {\alpha + s}, \quad
\mu = p(4) = \frac C {\alpha + 4}\,.
$$
Thus, the average total surface area and the average size are defined by the
formulas
$$
S(t) \sim \frac {CV}{(\alpha + 3)(\alpha + 4)}t\,, \eqno (14)
$$
$$
\lambda(t) \sim \frac {(\alpha + 1)(\alpha + 4)} C t^{-1} \eqno (15)
$$
and the average fragment number is
$$
N(t) \sim \frac {C^3 V}{(\alpha + 4)^3
(\alpha + 1)(\alpha + 2)(\alpha + 3)}t^3\,.
$$
Using (13), let us transform the equation (12),
$$
F(s+1) = \frac {\alpha + s}{\alpha + 1}F(s)\,.
$$
The solution of this functional equation which
satisfies to the normalization condition
$F(1,t) = 1$ is
$$
F(s) = \frac {\Gamma(\alpha + s)}{\Gamma(\alpha + 1) (\alpha + 1)^{s-1}}\,.
$$
Computing the inverse Mellin transformation, we find the limit distribution
on fragment size density
$$
f_\infty (r) = \frac
{(\alpha + 1)^{\alpha + 1} r^\alpha e^{-(\alpha + 1)r}}
{\Gamma(\alpha + 1)}\,.
$$
It is in accordance with the result of the work \cite{Our} which
has been obtained in the case when $\alpha$ is the integer.
However, the last distribution density differs essentially
from the logarithmically normal Kolmogorov law.

\end{document}